# SUMMARY OF FERMILAB'S RECYCLER ELECTRON COOLER OPERATION AND STUDIES*

L.R. Prost#, A. Shemyakin, FNAL, Batavia, IL 60510, U.S.A.

*Abstract*

Fermilab's Recycler ring [1] was used as a storage ring for accumulation and subsequent manipulations of 8 GeV antiprotons destined for the Tevatron collider. To satisfy these missions, a unique electron cooling system was designed, developed and successfully implemented [2]. The most important features that distinguish the Recycler cooler from other existing electron coolers are its relativistic energy, 4.3 MV combined with 0.1 – 0.5 A DC beam current, a weak continuous longitudinal magnetic field in the cooling section, 100 G, and lumped focusing elsewhere. With the termination of the Tevatron collider operation, so did the cooler. In this article, we summarize the experience of running this unique machine.

## COOLING PERFORMANCE

In the Recycler Electron Cooler, the beam is immersed in a longitudinal magnetic field at the gun and in the cooling section (CS); because the magnetic field is weak in the CS, 100 G, all estimations and results have been using a 'non-magnetized' cooling formalism [3].

### *Longitudinal cooling force*

The drag rate, $\dot{p}$, here measured by the 'voltage jump' method [4] (similar to [5]), represents the longitudinal cooling force averaged over all antiprotons. Hence, to interpret a drag rate as a cooling force experienced by the central particle, the antiproton beam needs to have a small rms momentum spread and a small transverse emittance. Early on, the dependence of the drag rate on the electron beam position offset with respect to the antiproton beam centroid trajectory lead to underestimating the actual cooling force. This is illustrated on Figure 1a, where the drag rate data are fitted with a simple expression of the cooling force (i.e. the so-called Binney formula e.g.: Ref. [6]) along with the cooling force reconstructed from incorporating the radial dependence of the drag rate. At the centre, the cooling force is higher than the measured drag rate by almost a factor of 2.

This effect, mostly due to the difficulty to control the transverse emittance, was creating a large scatter of the measured drag rates. Eventually, several adjustments to the procedure were made in order to maintain a low transverse emittance: the transverse stochastic cooling system was left on during the measurements; the antiproton beam was scraped down to the limit at which a reasonable resolution of the Schottky detector remained, $N_p \sim 1\times10^{10}$; and strongest cooling was applied between measurements.

Equally important was a decreased of the electron angles spread across the beam. These measures allowed improving the reproducibility of the results, and the antiproton beam transverse emittance measured with the flying wire, $\varepsilon_{n,95\%}$, was < 0.3 µm (normalized, 95%). An example of the drag rates obtained with the improved measurement procedure is shown in Figure 1b. In this case the measured drag rates and the longitudinal cooling force are within 8%.

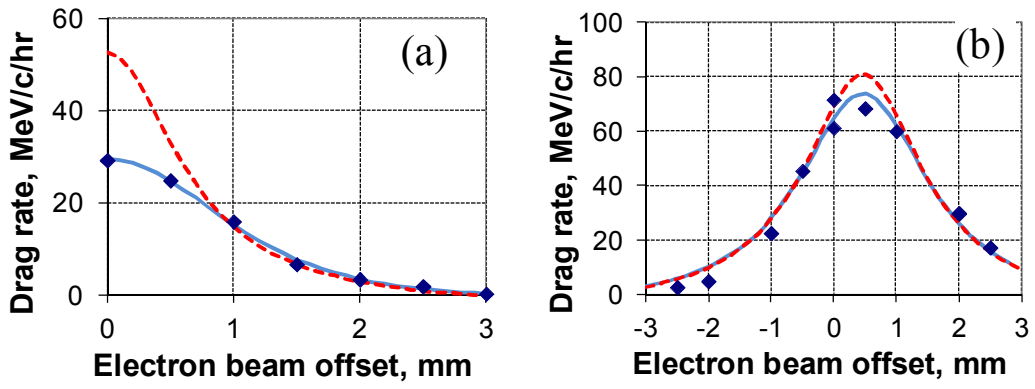

Figure 1: Drag rate as a function of the electron beam offset. The red curve is the reconstructed cooling force. Voltage jump ≡ 2 kV. (a) $I_e$ = 0.1 A, $N_p = 4\times10^{10}$, $\varepsilon_{n,95\%}$~0.5 µm, July 2007; (b) $I_e$ = 0.3 A, $N_p = 1.3\times10^{10}$, $\varepsilon_{n,95\%}$ < 0.3 µm, December 2010.

### *Electron beam quality and its optimization*

The beam quality for cooling can be characterized by the electron beam energy spread and angles in the cooling section.

The energy spread is mostly determined by the terminal voltage ripple. It primarily comes from the chain charging current fluctuations induced by the chain rotation and was estimated to be ~150 V based on beam trajectory measurements in a high dispersion region [7].

Drag rate measurements were the primary tool to optimize the beam angles. Figure 2 shows the dependence of drag rates on the beam current recorded over the years. The significant enhancement of the cooling force came mainly from three improvements that decreased the electron angles in the cooling section.

First, focusing was optimized by adjusting the corrector quadrupoles from drag rate measurements at the electron beam periphery [8].

Second, a beam-based procedure for aligning the magnetic field in the cooling section was developed [9]. The displacement of ten CS's individual modules with respect to one another due to the ground motion effectively introduces an undesirable transverse component to the field, which needed to be compensated at regular intervals (~twice a year) to preserve optimum cooling.

Finally, the electron angles were found to be affected by ions created by the electron beam and captured by its own space charge. While there were many ion clearing electrodes along the beam line, the remaining ion neutralization ~2% still significantly affected focusing for

# lprost@fnal.gov

beam currents $\gtrsim$ 100 mA. The remedy to decrease the average ion concentration was found to be periodic interruptions of the electron beam (2 μs with a frequency up to $f_{int}$ = 100 Hz depending on the beam current; so-called ion clearing mode) [10].

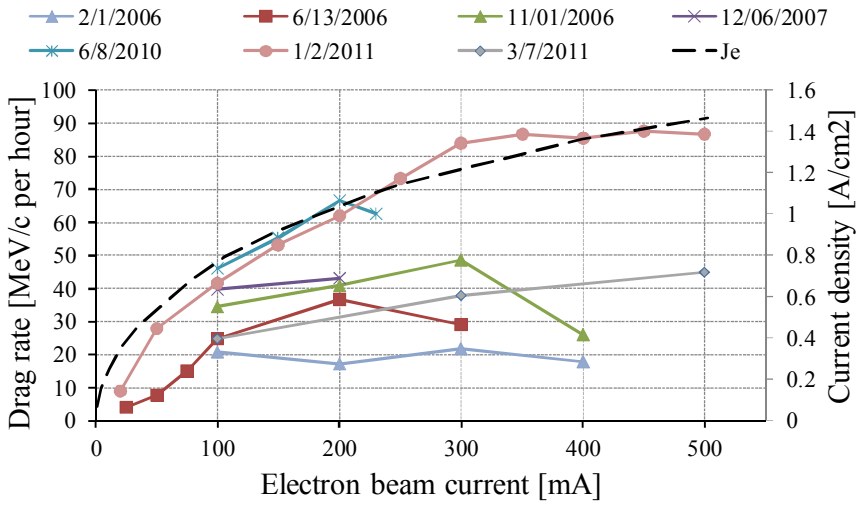

Figure 2: Drag rate as a function of the beam current measured on axis at various dates with a 2 kV voltage jump. The current density calculated at the beam centre (dashed curve) is shown for comparison.

For a 100 mA electron beam, the nominal for operation, including all sources of angles we estimate that the 1D, rms angle was ~0.1 mrad when the electron beam was fully optimized. Note that the flattening of the best curve on Figure 2 (labelled '1/2/2011'), at about 80 MeV/c per hour, is, at least partly, the result of the measurement procedure being inadequate for large drag rates.

## *Cooling rate*

While drag rate measurements were carried out to study and optimize the electron beam characteristics, the cooling rate defined as the difference of the time derivative of the momentum/emittances with the cooling system on and off, assesses numerically its actual effectiveness for operational conditions. The standard measurement procedure is described in Ref. [11].

Figure 3a summarizes electron cooling rates between 2006 and 2010. Over that time, the cooling rate for a given transverse emittance significantly increased due to the improvements to the electron beam quality discussed previously and highlighted on the plot. The arrows indicate the observed rate increase resulting from each of these beam optimization steps.

## *Cooling force vs. cooling rate*

The consistency between the two types of measurements can be checked by calculating the cooling rate expected from the cooling force model, which includes the radial dependence of the drag rates. The results are shown in Figure 3b (dash-dot pink curve) and compared with the subset of data from Figure 3a, which were measured at similar conditions. While this approach still slightly overestimates the cooling rate, it catches well its dependence on the antiprotons transverse emittance. Therefore, the drag rate and cooling rate measurement procedures give consistent descriptions of cooling properties. However, the large drag rate achieved for high beam currents did not correspondingly increase the cooling rate. Some data indicate that it could be due to non-linearities in the focusing solenoids preceding the cooling section [12], in which the beam size increases with the beam current, therefore increasing their effect on the electron beam angles.

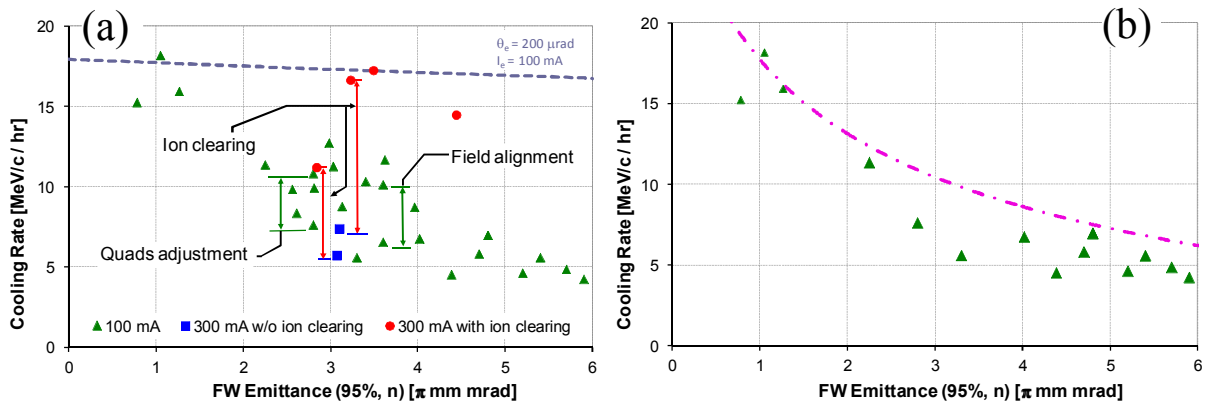

Figure 3: Longitudinal cooling rates (negated); (a) in 2006-2010; (b) Subset of the data where $I_e$ = 0.1 A, and for which the electron beam characteristics are similar. The dashed lines represent the expected cooling rates from models without (a) and with (b) inclusion of the radial dependence of the cooling force.

# OPERATION

In all previous electron coolers, electron and antiproton beams were overlapping concentrically. In the Recycler cooler, this configuration, which yields the maximum cooling rate, was not always required and induced a strong deterioration of the antiprotons lifetime. The solution to alleviate the latter issue was to displace the electron beam trajectory with respect to the antiproton beam orbit and adjust this offset to obtain the needed cooling. Typically, strongest cooling was applied only when preparing the antiprotons for extraction to the Tevatron.

One operational difficulty was energy drifts [7]. Keeping the equipment temperatures as constant as possible was found to be critical. Besides implementing better temperature regulation, one of the solutions was to rely on the displacement of the beam in a high dispersion region to measure the energy, and feed it back into the controls system. Figure 4a shows the beam voltage variation calculated from the beam displacement in the high dispersion region as a function of the Pelletron's temperature, when turning on after a shutdown. The most reliable indication of an energy mismatch was the shape of the Schottky momentum distribution, which becomes flat near its maximum. Hence, the parameters of the energy regulation loop were periodically corrected by making the momentum distribution as peaky as possible (Figure 4b).

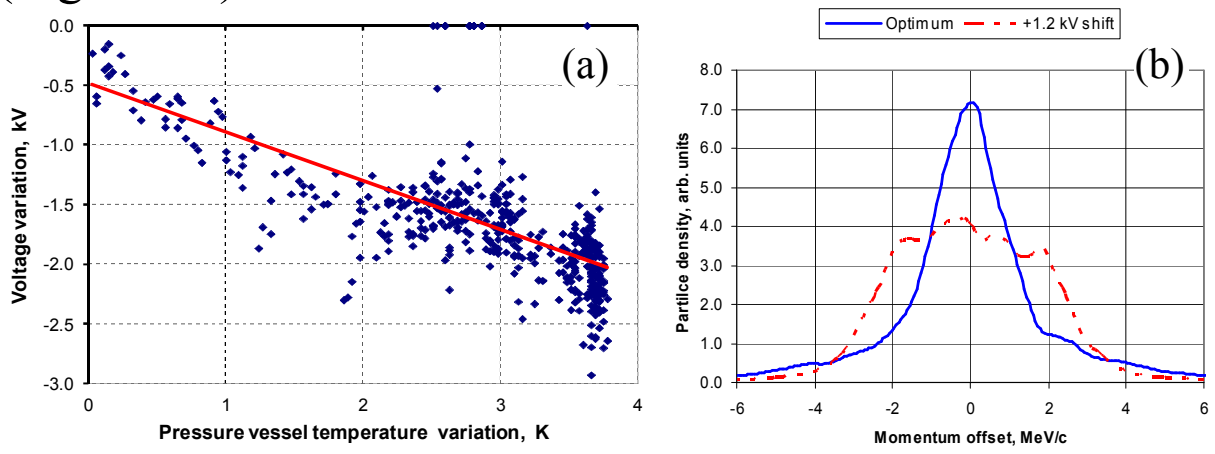

Figure 4: (a) Beam voltage variation vs. tank temperature (slope: -0.4 kV·K$^{-1}$); (b) Momentum distributions of the antiproton beam. Red line: energy shifted by 1.2 keV with respect to optimum; Blue line: optimum tuning of the electron beam energy.

When electron cooling was fully optimized and the ion clearing mode operational, the ability to apply strong cooling revealed two expected limitations: a transverse instability of the antiprotons with very small emittances and lifetime deterioration.

An impedance-driven beam instability [13] was foreseen and transverse dampers were designed and implemented greatly extending the stability region during accumulation. However, the extraction process includes complicated manipulations in the longitudinal phase space, and instabilities were observed a few times [14 and references therein].

In the Recycler, where antiprotons are typically accumulated for ~15 hours, preserving the antiproton beam lifetime is crucial. Although, we have not found a single parameter or combination of parameters that would uniquely determine the lifetime, it was observed, for instance, that the lifetime would increase quite significantly when increasing the bunch length.

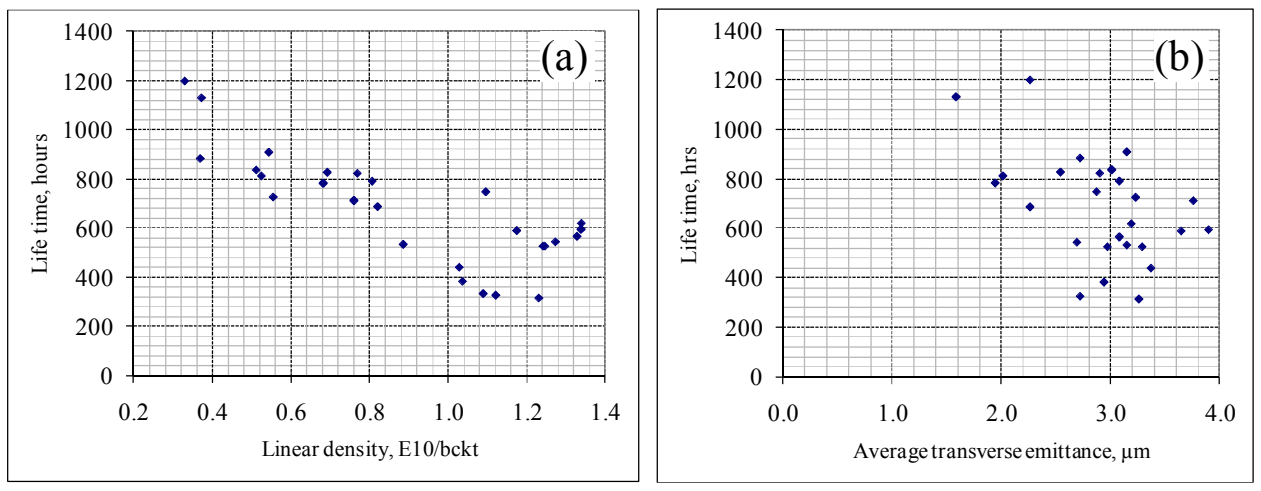

Figure 5: Beam lifetime in a steady state as a function of (a) the linear density and (b) the average Schottky emittance (95%, n). The linear density is calculated as a ratio of the total number of antiprotons, $\times 10^{10}$, to the length of the RF barriers gap, in units of 53-MHz buckets - the Recycler perimeter in this unit is 588 buckets.

Correspondingly, the data shown on Figure 5 seems to favour the linear density (Fig. 5a) rather than the transverse emittance (Fig. 5b) as the beam parameter most likely to correlate with the value of the beam lifetime.

In addition, while applying strong electron cooling deteriorates the lifetime, stochastic cooling improves it. A possible interpretation is that stochastic cooling acts on the far tail particles that electron cooling induces (probably similar to what is known as 'electron heating' for low-energy coolers [15]). Thus, from an operation point of view, it was very important to keep the stochastic cooling system properly tuned, even though its effect on the measured emittance of large stacks was insignificant.

## FINAL PERFORMANCE

Ultimately, the Recycler performance is characterized by its ability to store antiprotons efficiently and deliver bunches with adequate beam parameters to the Main Injector/Tevatron. In order to quantify the efficiency of the Recycler as a repository of antiprotons overall, a storage efficiency can be defined [16]. It includes injection and extraction efficiencies from and to the Main Injector, losses due to the antiprotons lifetime and accidental losses (e.g.: correctors' power supply trip, vacuum burst, and instability). For a typical accumulation and extraction cycle, where there is no accidental loss or operational issue, the storage efficiency was ~93%. Out of the 7% of beam which is lost, ~4% is due to injection and extraction inefficiencies while ~3% come from the antiprotons lifetime. At the same time, the Recycler was able to consistently cool the antiprotons to the adequate emittances (typically, 70 eV·s and 3 µrad, 95%, normalized), and deliver them to the Main Injector without deteriorating the quality of the bunches.